\documentclass[12pt]{iopart}
\usepackage{iopams}  
\usepackage{graphicx}  

\begin{document}
\title[Density dependence of the forbidden lines in Ni-like W]
{Density dependence of the forbidden lines in Ni-like tungsten}

\author{Yuri Ralchenko}

\address{Atomic Physics Division, 
National Institute of Standards and Technology, Gaithersburg, MD
20899-8422, USA}
\ead{yuri.ralchenko@nist.gov}

\begin{abstract}

The magnetic-octupole (M3) and electric-quadrupole (E2) transitions between the
ground state $3d^{10}~^1S_0$ and  the lowest excited $3d^94s$ $(5/2,1/2)$ J = 3
and J = 2 states in the Ni-like tungsten are shown to exhibit a strong
dependence on electron density $N_e$ in the range of values typical for tokamak
plasmas. Remarkably, the total intensity of these overlapping lines remains
almost constant, which may explain the strong emission in the 7.93~\AA~line
observed in tokamak experiments (Neu R et al. 1997 \JPB  {\bf 30} 5057).
Utilization of the M3 and E2 line ratios for density diagnostics in
high-spectral-resolution experiments is discussed as well.

\end{abstract}
\submitto{\jpb}
\maketitle

Since tungsten is considered to be a strong candidate for one of the
plasma-facing components in the next-generation tokamaks, the x-ray spectra from
its highly-charged ions are being actively studied in fusion devices, e.g. ASDEX
Upgrade tokamak \cite{Neu,Neu1}, and in electron beam ion traps (EBIT) 
\cite{PRA06,Neill}. The measured spectra are used to infer diverse and
substantial information on plasma parameters and to test advanced atomic
structure theories and collisional-radiative models.

The forbidden radiative transitions from highly-ionized tungsten are routinely
observed in x-ray and extreme ultraviolet (EUV) spectral regions
\cite{EUV,put,LLNL}. The Einstein coefficients for forbidden lines strongly
depend on the ion spectroscopic charge $Z_{sp}$, so that for 40 to 50 times
ionized tungsten atoms the electric-quadrupole (E2), magnetic-dipole (M1) and
even magnetic-octupole (M3) transition probabilities are sufficiently strong to
overcome collisional quenching in low-density plasmas. Since the intensities of
the forbidden lines are sensitive to the balance of radiative and collisional
processes, they often serve as an important diagnostic tool in fusion,
astrophysical and laboratory plasmas (see, e.g., \cite{Griem}).

\begin{figure}
\centering
\includegraphics[width=1\textwidth]{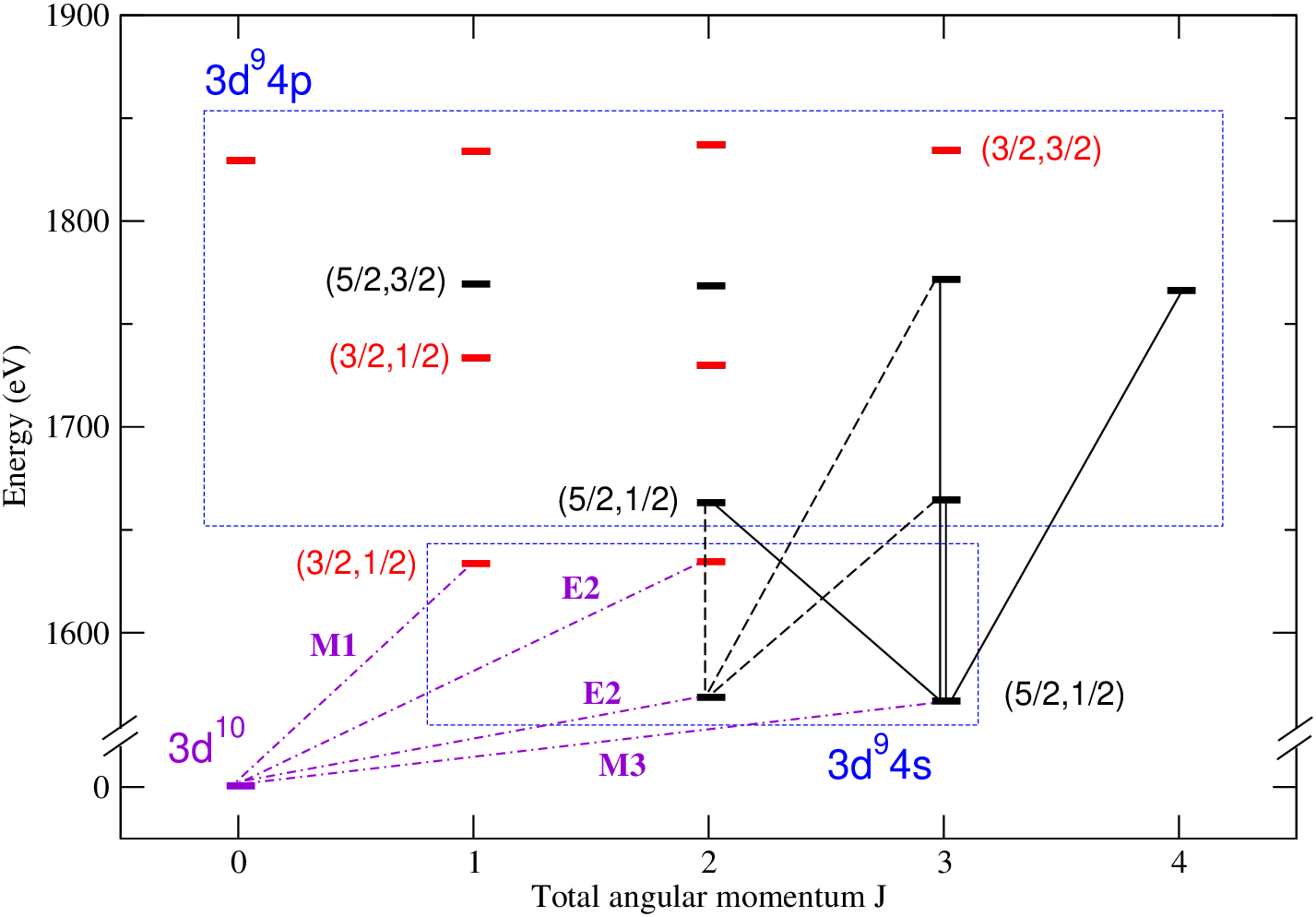}

\caption{\label{fig1}(Colour online.) 
Energy scheme of the $3d^{10}$, $3d^94s$, and 
$3d^94p$ configurations in Ni-like
tungsten. The forbidden $3d^{10} - 3d^94s$ transitions are shown by dot-dashed
lines with transition types indicated next to lines. The excited levels with the total angular momentum of the $3d^9$ core 
$J_c$ = 5/2 are shown in black, and the $J_c$ = 3/2 levels are shown in red.
The dominating excitation channels from the $(5/2,1/2)_3$ level are
shown by solid lines, and the E1 radiative transitions into the $(5/2,1/2)_2$
level are shown by dashed lines.
}
\end{figure}

One of the most prominent lines observed in the x-ray spectra of the
highly-charged tungsten  is a spectral line at 7.93~\AA\footnote{The wavelength
determined from the tokamak spectra \cite{Neu,Neu1} was 7.94~\AA, while our
recent EBIT measurements \cite{PRA06} gave a slightly smaller value of 7.93~\AA.
For consistency, it is the latter wavelength that is being used throughout this
paper.},~which originates from the Ni-like ion W {\sc XLVII}. (See, for
instance, Fig. 4 of Ref. \cite{Neu1} and Fig. 1 of Ref. \cite{PRA06}.) This line
is, in fact, an overlap of two forbidden lines, namely, the M3 line
$3d^{10~1}S_{0} - 3d^94s~(5/2,1/2)_3$ and the E2 line $3d^{10~1}S_{0} -
3d^94s~(5/2,1/2)_2$ with theoretical wavelengths of about 7.94~\AA~and 7.93~\AA,
respectively, as confirmed by several independent calculations
\cite{PRA06,Fournier,Safr}. In Ref. \cite{PRA06} we showed that in order to
correctly calculate the intensity of the 7.93~\AA~line in a low-density plasma
of EBIT, one has to accurately take into account both M3 and E2 transitions. We
also discussed the identification and population mechanisms for all four
forbidden lines between the first excited configuration $3d^94s$ and the ground
state $3d^{10~1}S_0$. These transitions (see Table \ref{Table1}) are indicated
by dot-dashed lines in Fig. \ref{fig1}, which presents the energy structure of
the $3d^{10}$, $3d^94s$, and $3d^94p$ configurations in the Ni-like tungsten.
For such a highly-charged heavy ion, jj-coupling is the most appropriate
coupling scheme, which is confirmed by the level grouping into jj-terms 
(Fig.~\ref{fig1}).

\begin{table}
\caption{\label{Table1}Calculated wavelengths (in \AA) and transition 
probabilities (in s$^{-1}$) for forbidden transitions 
$3d^{10}~^1S_0 - 3d^94s$ in Ni-like
tungsten. Notation a[b] denotes a~$\times$~10$^b$.}
\begin{indented}
\item[]\begin{tabular}{@{}llcccccc}
\br
Type  & Upper level& \centre{3}{Wavelengths (\AA)} & \centre{3}{Transition probabilities (s$^{-1}$)}\\
&&\crule{3}&\crule{3}\\
&&Ref.\cite{PRA06}&Ref.\cite{Fournier}&Ref.\cite{Safr}&This work&Ref.\cite{Fournier}&Ref.\cite{Safr}\\
\mr
M3 & $(5/2,1/2)_3$& 7.940& 7.945& 7.938& 9.35[3]& -- & 8.22[3]\\
E2 & $(5/2,1/2)_2$& 7.930& 7.935& 7.929& 5.94[9]& 5.92[9]& 5.32[9]\\
M1 & $(3/2,1/2)_1$& 7.616& 7.620& 7.614& 1.37[4]& --& 1.63[4]\\
E2 & $(3/2,1/2)_2$& 7.610& 7.614& 7.608& 4.55[9]& 4.51[9]& 4.04[9]\\
\br
\end{tabular}
\end{indented}

\end{table}

It was recently pointed out \cite{put1} that the M3 line, with its small
transition probability of $A_{M3} \approx 9 \times 10^{3}$ s$^{-1}$, may be
collisionally quenched in tokamak plasmas with electron density N$_e$ $\sim$
$10^{14}$ cm$^{-3}$, which is about three orders of magnitude higher than that
in an EBIT. To address this problem, we calculate here the intensities of the
$3d^{10} - 3d^94s$ forbidden lines for a wide range of electron densities $N_e$
from $10^{11}$ cm$^{-3}$ to $10^{15}$ cm$^{-3}$ and electron temperatures $T_e$
from 1000 eV to 5000 eV. This span of densities and temperatures covers the
typical values in tokamaks \cite{Neu,Neu1}. The line intensities are calculated
using the collisional-radiative code {\sc NOMAD} \cite{NOMAD} and the
relativistic atomic structure and collision code {\sc FAC} \cite{FAC}. The
details of our modelling are described elsewhere \cite{PRA06}, the only
difference being the use of a Maxwellian electron energy distribution function
for the thermal tokamak plasma discussed here. The simulations were performed in
the steady-state approximation. Although here we included six ionization stages
from W {\sc XLV} to W {\sc L} with total of more than 2400 levels, the main
conclusions can be derived considering only the levels within the Ni-like W {\sc
XLVII}.

\begin{figure}
\centering
\includegraphics[width=1\textwidth]{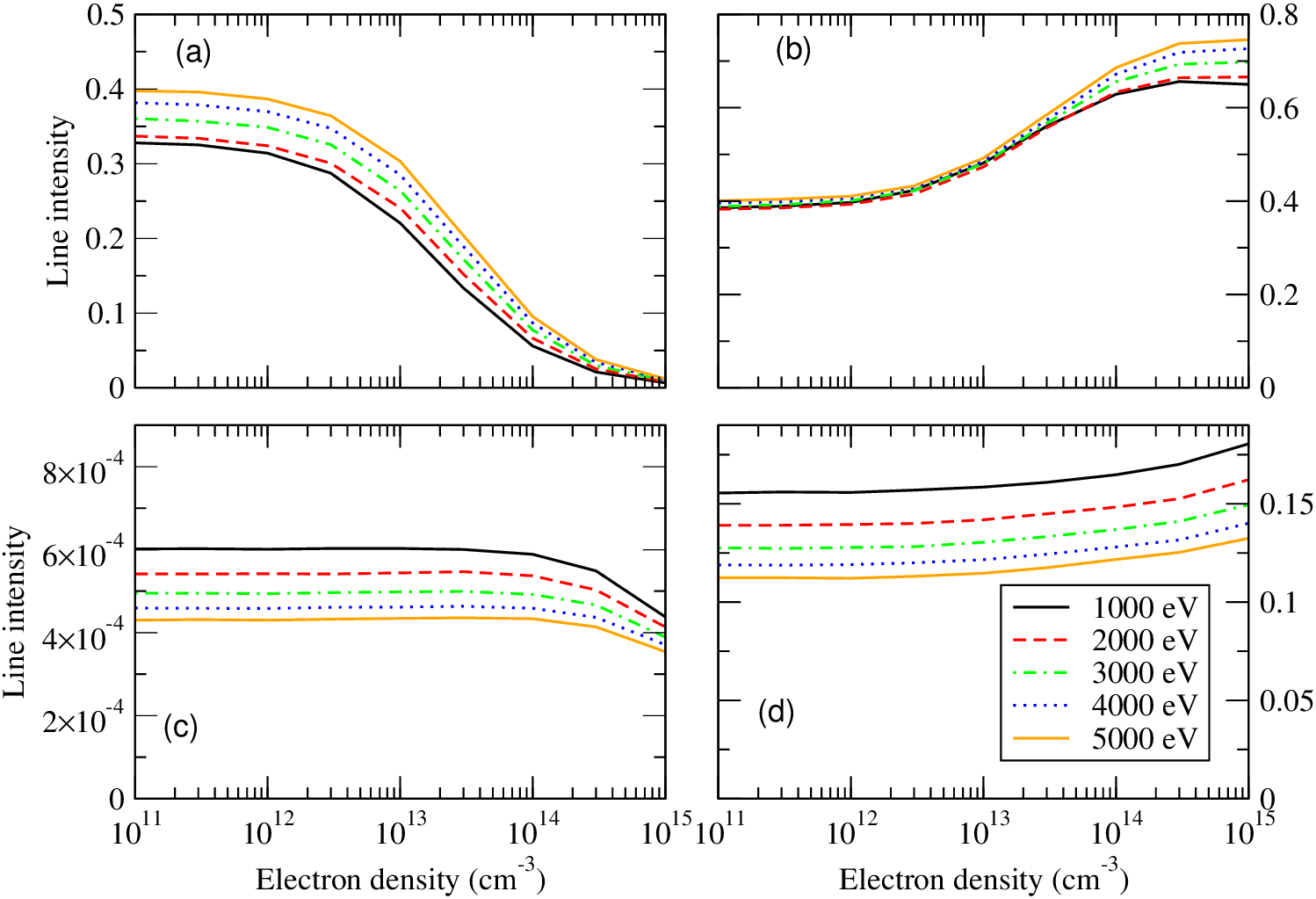}

\caption{\label{fig2}(Colour online.) 
Calculated intensities (relative to the E1 $3d^{10}~^1S_0 - 3d^94f
(3/2,5/2)_1$ line intensity) for the forbidden lines $3d^{10} -
3d^94s$ as a function of electron density for $T_e$ = (1000 -
5000) eV: (a) M3 line $^1S_0 - (5/2,1/2)_3$, (b)
E2 line $^1S_0 - (5/2,1/2)_2$, (c) M1 line $^1S_0
- (3/2,1/2)_1$, (d) E2 line $^1S_0 - (3/2,1/2)_2$.
} 
\end{figure}

\begin{figure}
\centering

\includegraphics[width=0.8\textwidth]{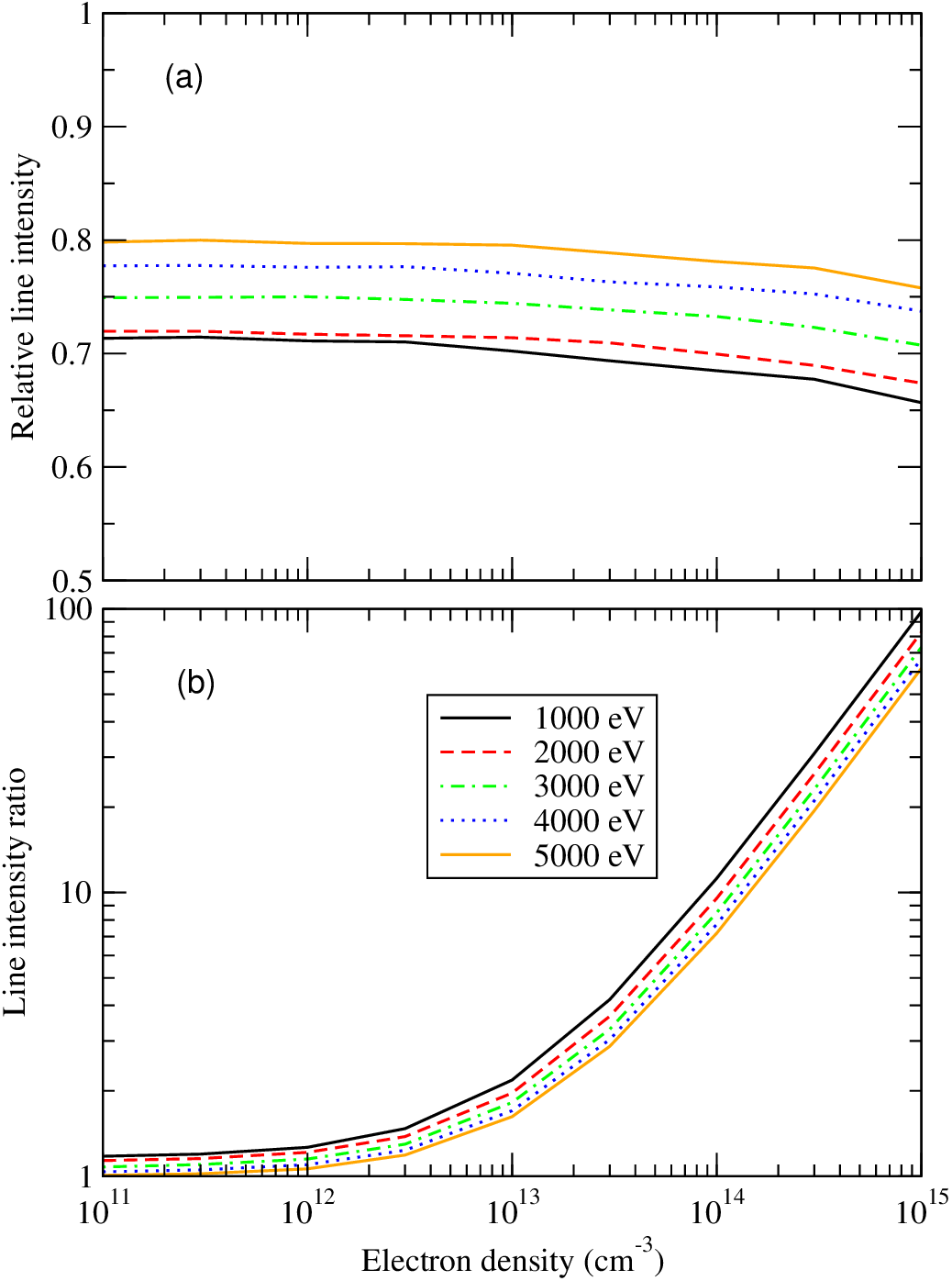}

\caption{\label{fig3}(Colour online.) 
Calculated line intensities for $T_e$ = (1000 -
5000) eV: (a) Sum of the relative intensities
for the M3 line $^1S_0 - (5/2,1/2)_3$ and E2 line $^1S_0 - (5/2,1/2)_2$;
(b) Ratio of line intensities $I_{E2}/I_{M3}$.
}
\end{figure}

It is convenient to present the results in terms of the line intensities
relative to the strongest dipole-allowed (E1) line in the Ni-like ion, namely,
the $3d^{10~ 1}S_0 - 3d^94f~(3/2,5/2)_1$ transition at 5.689~\AA. These
intensity ratios for the four $3d^{10} - 3d^94s$ forbidden lines are shown in
Fig. \ref{fig2}.  At lowest densities, the relative intensity for the weak M3
line $^1S_0 - (5/2,1/2)_3$ (Fig.\ref{fig2}(a)) is seen to remain approximately
constant up to N$_e$ $\approx$ $10^{12}$ cm$^{-3}$. For higher electron
densities, it indeed begins to decrease rapidly due to the collisional quenching
of the upper level. However, the relative intensity for the strong  E2 line
$^1S_0 - (5/2,1/2)_2$ with transition probability of $A_{E2} \approx$ $6 \times
10^9$ s$^{-1}$ {\it increases} with density beginning from the same value of
10$^{12}$ cm$^{-3}$ (see Fig.\ref{fig2}(b)). As these two lines closely overlap,
it is their total intensity that has been measured in the tokamak and EBIT
experiments. The total relative intensity, presented in Fig. \ref{fig3}(a),
remains approximately {\it constant} over the whole range of densities from
10$^{11}$ cm$^{-3}$ to 10$^{15}$ cm$^{-3}$. For instance, at the electron
temperature of 4000 eV, which is close to experimentally measured values
\cite{Neu,Neu1}, the intensity changes from about 0.78 at the lowest density to
0.76 at $3 \times 10^{13}$ cm$^{-3}$ and to 0.73 at $10^{15}$ cm$^{-3}$. This
interplay between the M3 and E2 line intensities is certainly not accidental. 

For low densities, the primary depopulation channel for the $3d^94s$
$(5/2,1/2)_3$ level is the M3 radiative decay: at $N_e$ = $10^{11}$ cm$^{-3}$ it
is almost two orders of magnitude stronger than collisional depopulation. For
higher densities, however, electron-impact collisions become more important so
that at the typical tokamak value of $N_e$ = $3 \times 10^{13}$ cm$^{-3}$, the
collisional excitation rate from the $3d^94s$ $(5/2,1/2)_3$ level is 4 times
larger than the M3 transition probability. It is well known that the collisional
excitation would preferentially go into the nearest levels that can be excited
via dipole-allowed collisions, that is, the levels of the $3s^94p$ configuration
(see Fig. \ref{fig1}). Moreover, the collisional rates are the strongest for
those transitions that do not result in the rearrangement of the core $3d^9$.
Therefore, in terms of jj-coupling, the excitation from the $3d^94s$
$(5/2,1/2)_3$ level with the total angular momentum of the core $J_c$ = 5/2
would primarily proceed into the $(5/2,1/2)$ or $(5/2,3/2)$ terms of the
$3d^94p$ configuration.

It is also possible to determine which specific levels within those terms would
be mainly populated via collisions. The selection rules for the dipole-allowed
electron-impact excitation \cite{exc} indicate that for the $s-p$ transitions,
the final levels should have total angular momentum $J_f$ differing by not more
than one unit from the initial value $J_i$. Therefore, the excitation from the
$3d^94s$ $(5/2,1/2)_3$ level would predominantly populate the $3d^94p$ levels
with $J_f$ = 2, 3, and 4. Indeed, our simulations show that at $N_e$ = $3 \times
10^{13}$ cm$^{-3}$ and $T_e$ = 4000 eV this excitation channel amounts to more
than 80~\% of the total collisional population outflux from this level.

The next important step in determination of the population redistribution
channels can be made by considering the radiative decays from the $3d^94p$
levels. Due to the $|\Delta J| \leq 1$ selection rule, the $J \geq 2$ levels of
the $3d^94p$ configuration do not have allowed electric-dipole transitions into
the ground state $3d^{10}~^1S_0$. On the other hand, the $J_f$ = 2 and 3 levels
with $J_c$ = 5/2 have strong ($A \approx 10^{10}-10^{11}$ s$^{-1}$) E1 decays
into the $3d^94s~(5/2,1/2)_2$ level, which is the upper level of the E2 line.
(The $3d^94p$ $(5/2,3/2)_4$ level decays radiatively back into the $3d^94s$
$(5/2,1/2)_3$.) Since the electric-dipole radiative decays of the $3d^94p$
levels remain the dominant depopulation channel over a wide range of densities,
at least up to $N_e \sim 10^{21}$ cm$^{-3}$, this in turn means that under 
tokamak conditions a significant part of the upward population flux from the
$3d^94s~(5/2,1/2)_3$ level would be redirected into the $3d^94s~(5/2,1/2)_2$
level followed by the E2 transition into the ground state. Hence, although the
collisional population redistribution between these two levels does modify the
M3 and E2 line intensities, their sum intensity, i.e., the total intensity of
the experimentally measured 7.93~\AA~line, remains almost constant over a large
range of densities.

Although this unresolved 7.93~\AA~line seems to be insensitive to $N_e$, the
intensity ratio of the M3 and E2 lines may become a very sensitive tool for
density diagnostics in tokamak plasmas, provided these two lines can be
resolved. Table \ref{Table1} shows that the calculated wavelength difference
$\Delta \lambda(E2-M3) $ is about 0.010~\AA, which is smaller than the
experimental resolving limit of 0.015~\AA~of Refs. \cite{Neu,Neu1}. The Doppler
width in a plasma of $T_e = 4000$ eV is approximately 0.003~\AA, so that a
spectrometer with resolving power of several thousands would be sufficient to
resolve the M3 and E2 lines. Figure \ref{fig3}(b) shows that the intensity ratio
$I_{E2}/I_{M3}$ monotonically increases from $\sim$1.7 at $N_e$ = $10^{13}$
cm$^{-3}$ to about 10 at $10^{14}$ cm$^{-3}$, and reaches almost 100 at
$10^{15}$ cm$^{-3}$. This range of values makes the $I_{E2}/I_{M3}$ ratio well
suited to density diagnostics in tokamaks.

One may ask whether other Ni-like ions might provide better opportunities for
the determination of the density-dependent $I_{E2}/I_{M3}$ ratio. Since the
$\Delta n = 0$ energy difference between the $(5/2,1/2)_3$ and $(5/2,1/2)_2$
levels of $3d^94s$ varies as  $Z_{sp}$ and the M3 and E2 transition energies are
proportional to $Z_{sp}^{2}$, the required resolution depends on the
spectroscopic charge as $\lambda/\Delta \lambda \propto Z_{sp}^{-1}$. Although
the fit of the results calculated with FAC in the range of $Z_{sp}$ = 29 -- 58
gives a slightly weaker dependence of $\lambda/\Delta\lambda \propto
Z_{sp}^{-0.8}$, it would seem likely that using the elements heavier than
tungsten would ease the spectral resolution requirements. However, this is not
the case, primarily due to  a very strong $Z_{sp}$-dependence, $A_{M3}$ $\propto
~ Z_{sp}^9$, for the M3 transition probability, which follows both from the
present {\sc FAC} calculations and from the results of Ref. \cite{Safr}. It is
obvious that the increasing $A_{M3}$ combined with the $1/Z_{sp}$ dependence of
the collisional $\Delta n = 0$ rates would shift the ratio sensitivity range
towards higher densities, outside the typical tokamak values. Using elements
lighter than tungsten, on the other hand, would drastically reduce the
probability of the M3 decay and thus enhance the collisional quenching so that
the M3 line would hardly be observed. It therefore seems quite peculiar that
tungsten and close elements may be the most suitable for such diagnostic
measurements.

Figure \ref{fig2}(c,d) also shows the density dependence of the two other
forbidden $3d^{10} - 3d^94s$ lines, namely, the M1 line $^1S_0 - (3/2,1/2)_1$
and the E2 line $^1S_0 - (3/2,1/2)_2$. The M1 line at 7.616~\AA~(Table
\ref{Table1}) is seen to be extremely weak, which is due to a strong quenching
M1 decay into the $(5/2,1/2)_2$ level of the same configuration $3d^94s$
\cite{PRA06,EUV}, and therefore can hardly be observed. The second E2 line at
7.610~\AA~that has a high transition probability of $A \approx 4.5 \times 10^9$
s$^{-1}$ exhibits a very weak dependence on $N_e$ (Fig. \ref{fig2}(d)),
increasing its relative intensity by only about 15 \% over the four orders of
magnitude change in $N_e$. Thus, these two lines cannot be reliably used for
density diagnostics in tokamak plasmas.

To summarize, we discussed here the density dependence of the intensities of the
forbidden $3d^{10} - 3d^94s$ lines in Ni-like tungsten. While the
magnetic-octupole and electric-quadrupole lines from the lowest excited
$(5/2,1/2)$ term do show a strong dependence on $N_e$ due to collisional
redistribution of population between the levels, the total relative intensity of
these overlapping lines does not change. This may explain the high intensity of
the 7.93~\AA~ line in the tokamak experiments. We also showed that, provided a
high spectral resolution ($\lambda / \Delta \lambda \gtrsim 2000$) is achieved,
the ratio of the E2 and M3 lines from the Ni-like tungsten can be used for
density diagnostics in tokamaks.

\ack

This work was supported in part by the Office of Fusion Energy Sciences of the
U.~S. Department of Energy.

\end{document}